# Structure of the eigenfrequencies parameter space for the system of dissipatively coupled oscillators

Yulia P. Emelianova[a,b*], Alexander P. Kuznetsov[b,c], Ludmila V. Turukina[b,c], Igor R. Sataev[c], and Nikolai Yu. Chernyshov[b]

[a] *Department of Electronics and Instrumentation, Saratov State Technical University, Polytechnicheskaya 77, Saratov 410054, Russian Federation*

[b] *Department of Nonlinear Processes, Saratov State University, Bolshaya Kazachya 112A, Saratov 410012, Russian Federation,
apkuz@rambler.ru (A.P. Kuznetsov), lvtur@rambler.ru (L.V. Turukina), sataevir@rambler.ru (I.R. Sataev), nick.chernyshov88@gmail.com (N. Chernyshov)*

[c] *Kotel'nikov's Institute of Radio-Engineering and Electronics of RAS, Saratov Branch, Zelenaya 38, Saratov, 410019, Russian Federation,*

**Abstract**

Structure of the eigenfrequencies parameter space for three and four dissipatively coupled van der Pol oscillators is discussed. Situations of different codimension relating to the configuration of the full synchronization area as well as a picture of different modes in its neighborhood are revealed. The organization of quasi-periodic areas of different dimensions is considered. The results for the phase model and for the original system are compared.

## 1. Introduction

Problems concerned with the interaction of self-oscillating systems are important for the nonlinear science and various applications. In the simplest case of two oscillators, presentation of the results on the parameter plane "difference between eigenfrequencies – coupling parameter" is traditional. Such an approach for Adler-type models or for a circle map produces a picture of Arnol'd tongues, immersed in the region of quasi-periodic oscillations [1,2]. An increase of the number of oscillators makes it possible to obtain illustrations of another type, namely, to explore structure of the oscillators' eigenfrequencies (more precisely, frequency detunings) parameter space. This problem is very complex, since it must include a description of the different order resonances, their configuration, relative position, as well as a picture of quasi-periodic regimes of different dimensions. This problem was attracting for a long time for physicists and mathematicians in different aspects, such as the assessment of the size of periodic resonances depending on the coupling value [3-5], generalization of the Fairy tree and continued fractions to the case of a larger number of frequencies [6-10], detection of local and nonlocal bifurcations [11-13], structure of multi-frequency areas [14], experiments with electronic oscillators [15,16]. These aspects are also interrelated with problems of the forced synchronization of a resonant limit cycle on the torus [17-18], with the dynamics of coupled quasi-periodic generators [19], etc. However, the overall picture is still not built. We discuss here the problem concerned with the structure of the fundamental resonance and its neighborhood for dissipatively coupled oscillators.

---

[*] Corresponding author at: Department of Electronics and Instrumentation, Saratov State Technical University, Polytechnicheskaya 77, Saratov 410054, Russian Federation.
*E-mail address*: yuliaem@gmail.com (Yu.P. Emelianova)

It should be noted that it is very important for the discussed range of problems to choose a physically motivated model. For example, the system from [11] is not applicable for these purposes. The reason is in selection of a model: it does not have a stable equilibrium, and therefore does not describe the possibility of the full synchronization of oscillators with the frequency ratio of 1:1:1. At the same time, many of the results of this work are important and significant. The picture obtained in [15] relates to the experimental study of a specific electronic circuit. Thus, it is important to choose a fairly universal, but physically based system. We will use van der Pol oscillator as such system.

It was shown still in 1980 in the book of P.S. Landa [1] that for three van der Pol oscillators in the phase approximation the full synchronization area corresponding to the frequency ratio of 1:1:1 has the form of a parallelogram. This result was obtained analytically and is quite universal due to the universality of the model being used. However, modern methods and approaches can significantly complement and develop it. In the present paper we answer the following questions:
- Which regimes are observed outside the "Landa's parallelogram"?
- What will change if we move from the phase model to the original system?
- What will happen if we increase the number of oscillators in the chain?

Note that quasi-periodic oscillations of different dimensions are possible in this system. For such systems approaches to the analysis of (quasi-periodic) bifurcations are still underdeveloped [21, 22]. Therefore, the main investigation tool for such modes is an analysis of the Lyapunov exponents.

## 2. Phase model for the three oscillators

The chain of three coupled van der Pol oscillators is described by the following equations:

$$\ddot{x} - (\lambda - x^2)\dot{x} + x + \mu(\dot{x} - \dot{y}) = 0,$$
$$\ddot{y} - (\lambda - y^2)\dot{y} + (1 + \Delta_1)y + \mu(\dot{y} - \dot{x}) + \mu(\dot{y} - \dot{z}) = 0, \quad (1)$$
$$\ddot{z} - (\lambda - z^2)\dot{z} + (1 + \Delta_2)z + \mu(\dot{z} - \dot{y}) = 0.$$

Here $\lambda$ is an excitation parameter in each independent oscillator and characterizes negative friction; $\Delta_1$ is the frequency detuning between the second and the first oscillators; $\Delta_2$ is the frequency detuning between the third and the first oscillators; $\mu$ is the coefficient of dissipative coupling. The frequency of the first oscillator is taken as unity.

If excitation parameter λ, frequency detunings and coupling parameter are small, the method of slowly varying amplitudes [1,2] can be applied for the analysis of the equations (1). In addition, if assume that the oscillators move along their stationary orbits, it is possible to obtain equations in the phase approximation [1,14]:

$$\dot{\theta} = -\frac{\Delta_1}{2} - \mu\sin\theta + \frac{\mu}{2}\sin\varphi,$$
$$\dot{\varphi} = \frac{\Delta_1 - \Delta_2}{2} - \mu\sin\varphi + \frac{\mu}{2}\sin\theta. \quad (2)$$

Here $\theta = \psi_1 - \psi_2$, $\varphi = \psi_2 - \psi_3$ are relative phases of the oscillators, and all the parameters are normalized by λ, so they may be not small.

Let us find the explicit boundaries of the "Landa's parallelogram". Full synchronization of all three oscillators corresponds to the stable equilibrium of the phase system (2). After setting $\dot{\theta} = 0$ and $\dot{\varphi} = 0$ in (2) we obtain

$$\sin\theta = -\frac{\Delta_1 + \Delta_2}{3\mu},$$
$$\sin\varphi = \frac{\Delta_1 - 2\Delta_2}{3\mu}. \quad (3)$$

Solutions (3) exist when the sinuses modulo less than unity. Therefore, the bifurcation conditions are of the form

$$\frac{\Delta_1 - 2\Delta_2}{3\mu} = \pm 1, \quad (4)$$

$$\frac{\Delta_1 + \Delta_2}{3\mu} = \pm 1. \quad (5)$$

The intersection of the lines (4) and (5) forms a parallelogram on the parameter plane $(\Delta_1, \Delta_2)$ as is shown in Fig. 1b. During the variation of the parameters, the solutions (3) appear in pairs: $(\theta_1, \theta_2)$ for the first equation and $(\varphi_1, \varphi_2)$ for the second equation. Thus, the system (2) has four equilibria, stable and unstable nodes and two saddles [14,17]. Crossing the borders of the parallelogram they approach each other in pairs ($\theta_1 \to \theta_2$ or $\varphi_1 \to \varphi_2$) and then merge and disappear simultaneously. Note that it leads from (4) and (5) that the synchronization area enlarges linearly with an increase of the coupling parameter µ.

Besides the full synchronization, two-frequency and three-frequency quasi-periodic regimes[1] are possible in the system (2). In the first case, the attractor of the phase model is an invariant curve, an example of which is shown in Fig. 2a. Such trajectories are conveniently classified by means of a kind of the rotation number $w = p:q$, which is determined by the number of significant intersections of the phase trajectory with the horizontal and vertical boundaries of the field $0 < \theta, \varphi < 2\pi$

---

[1] In these cases attractors for the original system are two-frequency and three-frequency tori.

[14,18,20]. So this number is $w$=1:3 for Fig. 2a. In the second case, the phase trajectories fill the whole "phase square", as is shown in Fig. 2b.

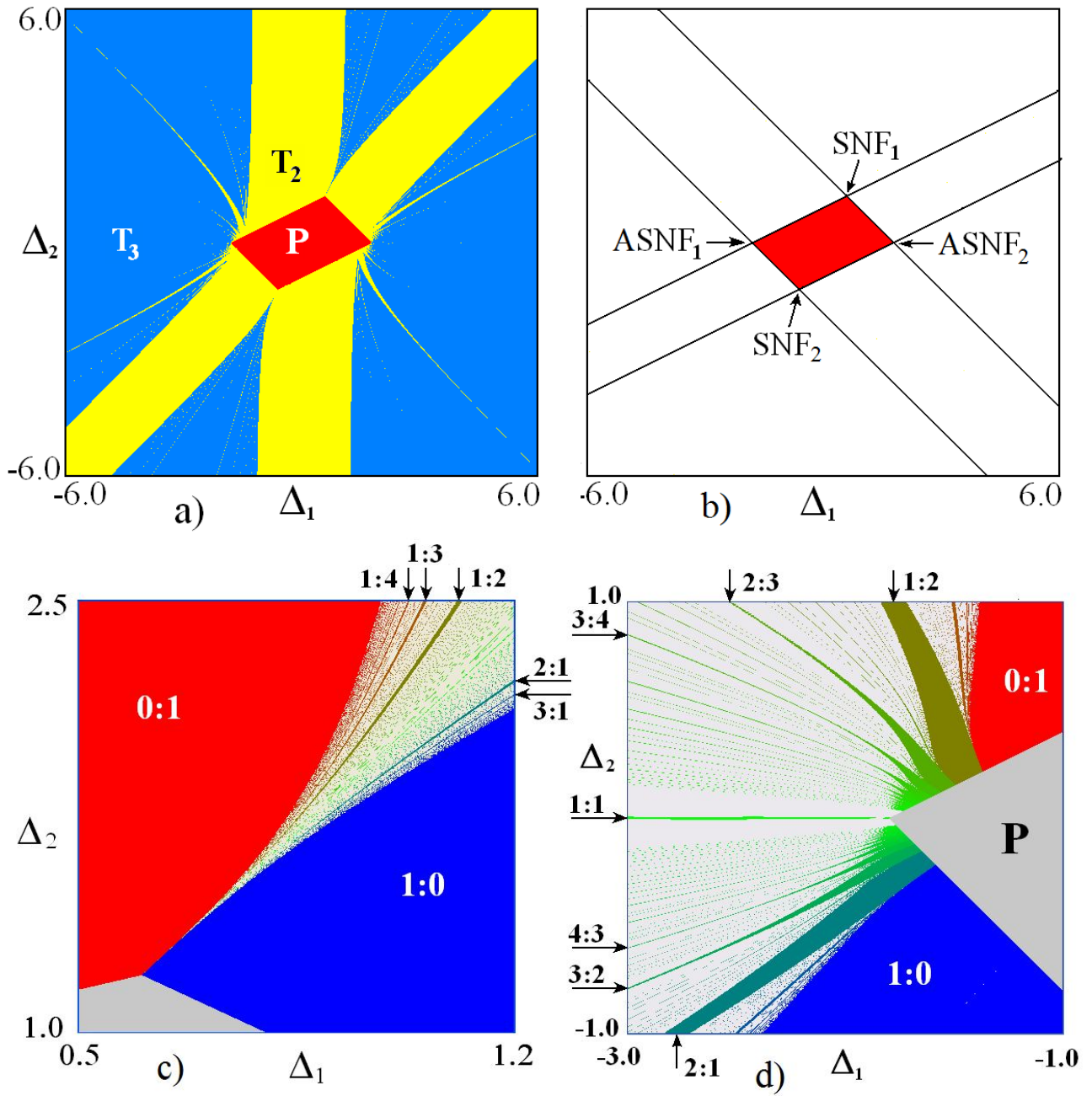

**Fig. 1.** (a) Chart of the Lyapunov exponents constructed for three dissipatively coupled phase oscillators on the frequencies parameter plane, (b) analytically obtained configuration of the full synchronization area, (c)-(d) charts of tori in the neighborhood of SNF and ASNF points. Value of coupling parameter is $\mu = 0.6$.

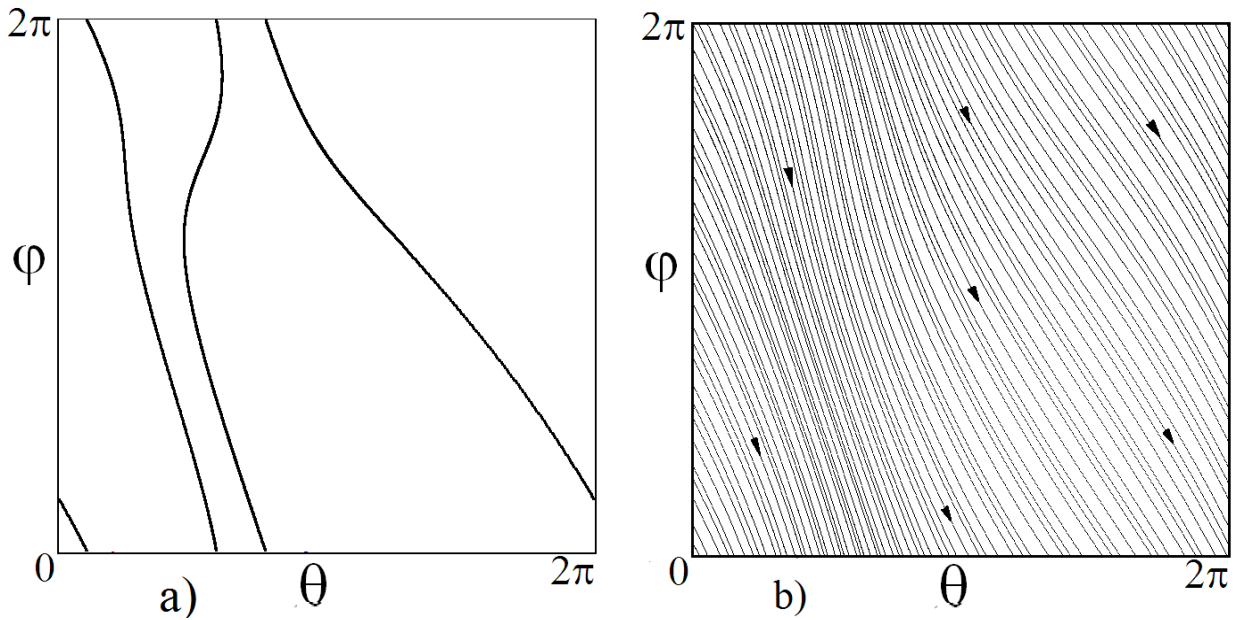

**Fig. 2.** Examples of phase portraits for the system (2). (a) Two-frequency resonance regime of the type 1:3 for $\Delta_1 = -1.5$, $\Delta_2 = 1$, $\mu = 0.6$; (b) three-frequency regime for $\Delta_1 = -1$, $\Delta_2 = 1$, $\mu = 0.25$.

Let us now define the character of regimes outside the parallelogram using the construction of the chart of Lyapunov exponents [14,18,20], which visualizes areas of the stable equilibrium $P$, two-frequency quasi-periodicity $T_2$ and three-frequency quasi-periodicity $T_3$. This chart is shown in Fig. 1a for the phase system (2).

One can see from Fig. 1a that the full synchronization area $P$ satisfies the analytical estimation and is bounded by the lines (4) and (5). The most representative two-frequency regimes in Fig. 1a look like two bars, at the intersection of which the full synchronization area is located. These bars correspond to the partial synchronization of two different pairs of oscillators, when their relative phase oscillates around some average value. They meet the resonance conditions $\Delta_1 \approx 0$ and $\Delta_1 \approx \Delta_2$. In terms of the original system the first of these two conditions means the coincidence of eigenfrequencies of the first and the second oscillators, and the second condition means the coincidence of the second and the third oscillators. Accordingly, the rotation numbers are equal to 0:1 and 1:0. The width of these bars is of the order of coupling parameter magnitude.

Intersection of the lines (4) and (5) corresponds to the codimension-two situations when all equilibriums come simultaneously into a one point on the phase plane. There are four such points on the boundary of the full synchronization area and they may be easily found by combining (4) and (5):

$$SNF_{1,2}: \Delta_1 = \mu, \Delta_2 = 2\mu;\ \Delta_1 = -\mu, \Delta_2 = -2\mu;$$
$$ASNF_{1,2}: \Delta_1 = 3\mu, \Delta_2 = 0;\ \Delta_1 = -3\mu, \Delta_2 = 0. \quad (6)$$

Areas of the two-frequency regimes have the most complex structure in the neighborhood of these points. In particular, there are many resonance regimes of different order. Their co-existence illustrate "charts of tori" in Fig. 1c and Fig. 1g. Different colors on these charts correspond to different rotation numbers $w$ [14,18,20].

For the first pair of SNF points a fan-shaped system of divergent areas of different order two-frequency resonance regimes is characteristic, with (at least on a visual level) a common peak directly at the point SNF. The neighborhood of the second pair of points – ASNF points – is more complicated. In this case, the two-frequency regime tongues do not decrease in size approaching the full synchronization area, but vice versa increase so that they have an extended contact with it, see Fig. 1d. Using the terminology of [11,12,16], the first pair of points (6) may be called as "saddle node fan", and the second pair of points may be associated with the point "accumulation of saddle node fans".

## 3. Three van der Pol oscillators

Let us refer to the original system (1). Figure 3 illustrates the corresponding chart of the Lyapunov exponents for small value of the control parameter $\lambda = 0.1$. Note that in accordance with the normalization by $\lambda$, all values of the parameters are ten times larger as compared with Fig. 1. We can see some features that are typical for the phase model. Namely, the form of the full synchronization area is close to a parallelogram, there are two basic bars of two-frequency regimes immersed in the area of three-frequency tori, there are fan-shaped structures of the two-frequency high order resonant tori. However, there are certain differences. The full synchronization area transforms from the parallelogram[2]. So instead of *SNF* points in Fig. 1 the characteristic vertices of a parallelogram are changed into the smooth lines. On the contrary, for *ASNF* points appreciably prolate vertices occur.

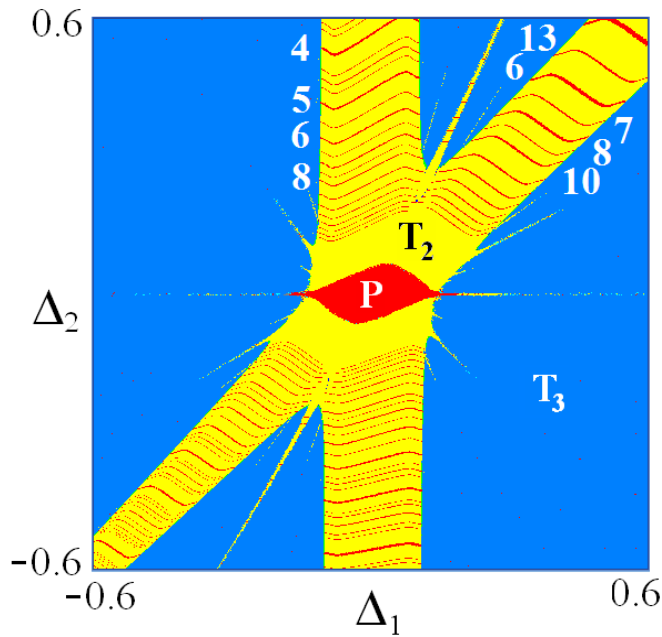

**Fig. 3.** Chart of the Lyapunov exponents for three coupled van der Pol oscillators on the frequency detunings parameter plane. Numbers correspond to cycle periods in the Poincaré section. Values of the parameters are $\lambda = 0.1$, $\mu = 0.04$.

---

[2] Note that a limit cycle corresponds to the full synchronization. Its disappearance corresponds to a saddle-node bifurcation in the Poincaré section.

The bifurcation analysis for the full synchronization area of the system (1) allows to see the differences in more details. The obtained results are shown in Fig. 4.

The first difference between the original system and the phase model is that saddle-node bifurcations of stable and unstable limit cycles do not occur simultaneously, but on different lines $SN_1$ and $SN_2$ that are shown in Fig. 4a by solid lines and dashed lines, respectively. So the "vertex" SNF of the parallelogram is changed into a smooth line. At the same time there are three cusp points for the bifurcation line of unstable regimes inside the full synchronization area.

Enlarged fragment in Fig. 4b shows that Neimark–Sacker bifurcation lines *NS* are involved in the destruction of ASNF points. These lines have common points (points R1 of resonance 1:1) with the saddle bifurcation lines. Thus, the highly prolate narrow areas of a full synchronization occurring instead of ASNF points are limited by the Neimark–Sacker bifurcation lines.

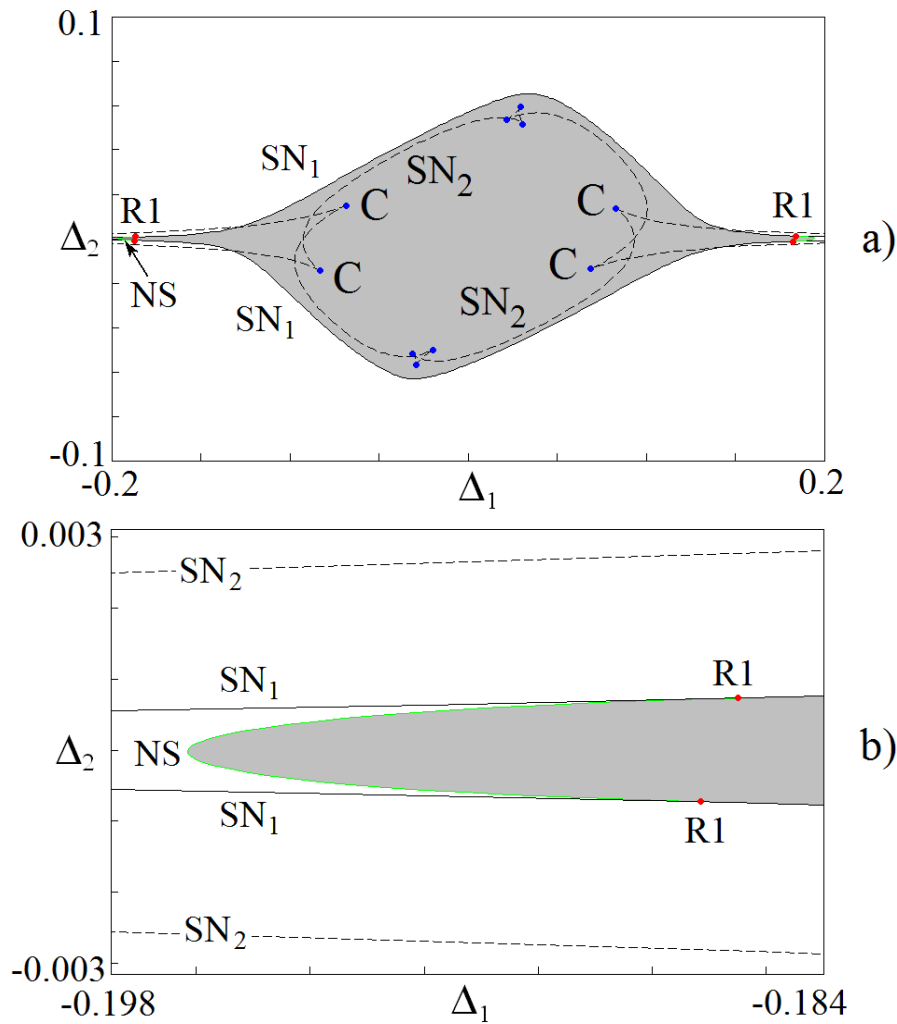

**Fig. 4.** Bifurcation lines for three coupled oscillators (1). Fig. b shows an enlarged fragment of Fig. a. Values of the parameters are $\lambda = 0.1$, $\mu = 0.04$. Solid lines identify bifurcations for stable regimes, and dashed lines – for unstable regimes. SN denote the saddle-node bifurcation lines ($SN_1$ corresponds to the merging of a stable node and a saddle, $SN_2$ corresponds to the merging of an unstable node and a saddle), C denote the cusp points, NS correspond to the Neimark–Sacker bifurcation lines, R1 denote resonances 1:1.

We can see resonance regions of higher order periodic regimes on the chart of Lyapunov exponents in Fig. 3 within the two main bars of two-frequency tori, which is not observed in the phase model. These regions correspond to different periods in the Poincaré section and are labeled by numbers in Fig. 3.

Figure 5 shows the chart of Lyapunov exponents for large value of the control parameter $\lambda = 1$. The phase approximation is not applicable in this case. Despite the large value of the control parameter, the full synchronization area keeps generally its form. However, there are significant differences. In addition to the synchronization area 1:1:1 there is another large full synchronization area. It corresponds to the harmonic resonance, when eigenfrequencies of the three oscillators are related as 1:3:3. Indeed, as follows from the definition of frequency detunings in (1), eigenfrequencies of the oscillators are $\omega_1 = 1, \omega_{2,3} = \sqrt{1 + \Delta_{1,2}}$, and the values $\Delta_1 = \Delta_2 = 8$ satisfy such relation between frequencies.

The Arnol'd resonance web structure [21,22] becomes visible for large values of the control parameter. This is a system of thin lines of two-frequency tori, on intersection of which the higher order periodic resonances arise. Some of these resonances are marked by numbers in Fig. 5.

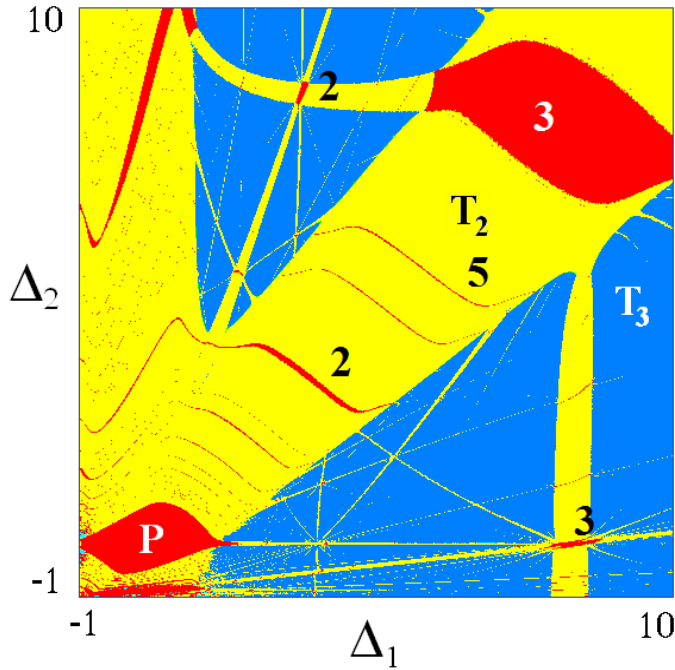

**Fig. 5.** Chart of the Lyapunov exponents for three coupled van der Pol oscillators on the frequency detunings parameter plane. Numbers correspond to cycle periods in the Poincaré section. Values of the parameters are $\lambda = 1, \mu = 0.4$.

## 4. Phase model for the four oscillators

Consider now the chain of four dissipatively coupled van der Pol oscillators:

$$\ddot{x} - (\lambda - x^2)\dot{x} + x + \mu(\dot{x} - \dot{y}) = 0,$$
$$\ddot{y} - (\lambda - y^2)\dot{y} + (1 + \Delta_1)y + \mu(\dot{y} - \dot{x}) + \mu(\dot{y} - \dot{z}) = 0,$$
$$\ddot{z} - (\lambda - z^2)\dot{z} + (1 + \Delta_2)z + \mu(\dot{z} - \dot{y}) + \mu(\dot{z} - \dot{w}) = 0, \quad (7)$$
$$\ddot{w} - (\lambda - w^2)\dot{w} + (1 + \Delta_3)w + \mu(\dot{w} - \dot{z}) = 0.$$

Here the meaning of the parameters is similar to those in equations (1). $\Delta_1, \Delta_2, \Delta_3$ denote eigenfrequency of the second, the third and the forth oscillator, respectively. The frequency of the first oscillator is taken as unity. Acting in a conventional manner, we obtain equations for the relative phases of oscillators in the phase approximation:

$$\dot{\theta} = -\frac{\Delta_1}{2} - \mu \sin\theta + \frac{\mu}{2}\sin\varphi,$$
$$\dot{\varphi} = \frac{\Delta_1 - \Delta_2}{2} + \frac{\mu}{2}\sin\theta - \mu\sin\varphi + \frac{\mu}{2}\sin\phi, \quad (8)$$
$$\dot{\phi} = \frac{\Delta_2 - \Delta_3}{2} + \frac{\mu}{2}\sin\varphi - \mu\sin\phi.$$

Conditions for the full synchronization are $\dot{\theta} = \dot{\varphi} = \dot{\phi} = 0$. After some simple transformations we can obtain from (8) equations for the sines of relative phases:

$$\sin\theta = -\frac{\Delta_1 + \Delta_2 + \Delta_3}{4\mu},$$
$$\sin\varphi = \frac{\Delta_1 - \Delta_2 - \Delta_3}{2\mu}, \quad (9)$$
$$\sin\phi = \frac{\Delta_1 + \Delta_2 - 3\Delta_3}{4\mu}.$$

During the variation of the parameters, solutions of the equations (9) appear in pairs: $(\theta_1, \theta_2)$, $(\varphi_1, \varphi_2)$ and $(\phi_1, \phi_2)$. Therefore, the system (8) has eight equilibrium states located at the vertices of a parallelepiped in the phase space $(\theta, \varphi, \phi)$. One of these equilibrium states is always stable, and the rest are saddles and unstable node.

If any of the three combinations of the parameters on the right side in (9) is varying, two faces of the parallelepiped approach each other and merge. All the eight pairs of fixed points merge and disappear simultaneously, as soon as the sine of one of the phase variables equals to unity. At the same time the steady state disappears and the full synchronization collapses. So we can obtain expressions for the saddle-node bifurcations of such type:

$$\mu = \pm\frac{\Delta_1 + \Delta_2 + \Delta_3}{4}, \quad (10)$$

$$\mu = \pm\frac{\Delta_1 - \Delta_2 - \Delta_3}{2}, \quad (11)$$

$$\mu = \pm \frac{\Delta_1 + \Delta_2 - 3\Delta_3}{4}. \qquad (12)$$

Thus, there are three variants of such bifurcation corresponding to the merging of the parallelepiped sides along one of the three phase axes.

Let us discuss the structure of the eigenfrequencies parameter space $(\Delta_1, \Delta_2, \Delta_3)$. Full synchronization area is defined by equations (10)–(12) that correspond to three pairs of planes whose intersection forms a polyhedron in the form of an *oblique parallelepiped*, Fig. 6. Its faces are surfaces of codimension-one bifurcations of the above three types, its edges are codimension-two bifurcations, and its verteces are codimension-three bifurcations.

It is convenient to represent sections of the polyhedron by planes $\Delta_2 = const$ and to consider corresponding parameter planes $(\Delta_1, \Delta_3)$, examples of which are shown in Fig. 7 and Fig. 8. Coupling parameter is assumed to be fixed $\mu = 0.4$.

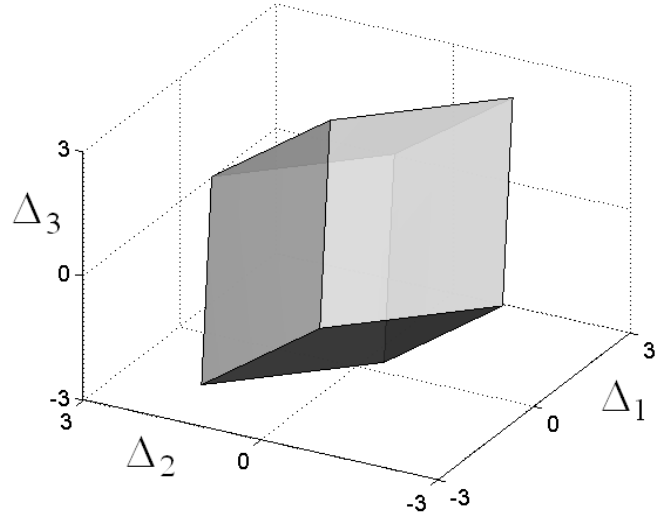

**Fig. 6.** Full synchronization area for the four phase oscillators on the frequency detunings parameter space $(\Delta_1, \Delta_2, \Delta_3)$.

Figure 7 shows the chart of Lyapunov exponents constructed for small value of the frequency $\Delta_2 = 0.2$. Figure 7b represents three pairs of lines given by equations (10)–(11). Their intersection forms a *hexagon*, which corresponds to the full synchronization area and is visible also on the chart of Lyapunov exponents. Thus, there are six codimension-two points which are vertices of the hexagon. Note that four-frequency regimes $T_4$ are possible now.

Chart of tori in the neighborhood of two vertices of the hexagon is presented in Fig. 7c. Near the upper "corner" of the full synchronization area in Fig. 7c we can see regions with the rotation numbers p: q: 0. It means that the third and the fourth oscillators are captured mutually. So the considered system is partly similar to the system of three coupled oscillators where the role of the third oscillator plays the captured pair of the third and the fourth oscillators. But these regions do not form "fan" structures: two vertices of the hexagon in Fig. 7c are closely spaced enough, and characteristic for the three oscillators picture which is shown in Figs. 3c,d is not observed in the pure state.

Straight lines bounding the full synchronization area in Fig. 7b are shifting with an increase of the frequency parameter $\Delta_2$ in accordance with the equations (10)–(11). Two pairs of lines are shifting downwards and one pairs of lines is shifting upwards. It is easy to show analytically that vertices of the hexagon merge in pairs simultaneously when $\Delta_2 = \mu$, so the shape of the full synchronization area looks like a triangle. Figure 8a shows the chart of Lyapunov exponents constructed for such a threshold situation.

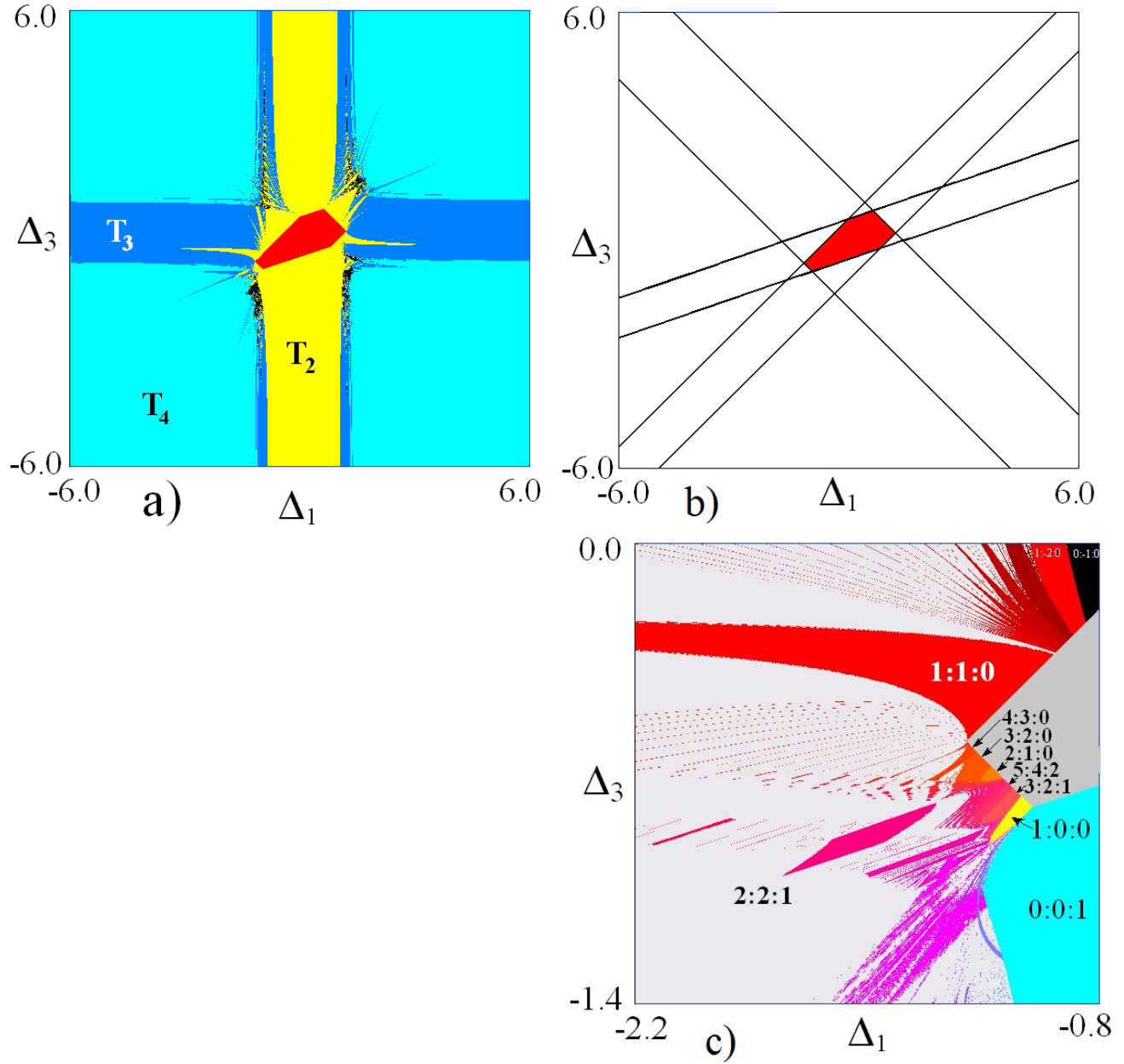

**Fig. 7.** (a) Chart of the Lyapunov exponents for the four dissipatively coupled phase oscillators on the frequency detunings parameter plane $(\Delta_1, \Delta_3)$; (b) analytically obtained configuration of the full synchronization area; (c) chart of tori in the neighborhood of two vertices of the full synchronization area. Values of the parameters are $\mu = 0.4$, $\Delta_2 = 0.2$.

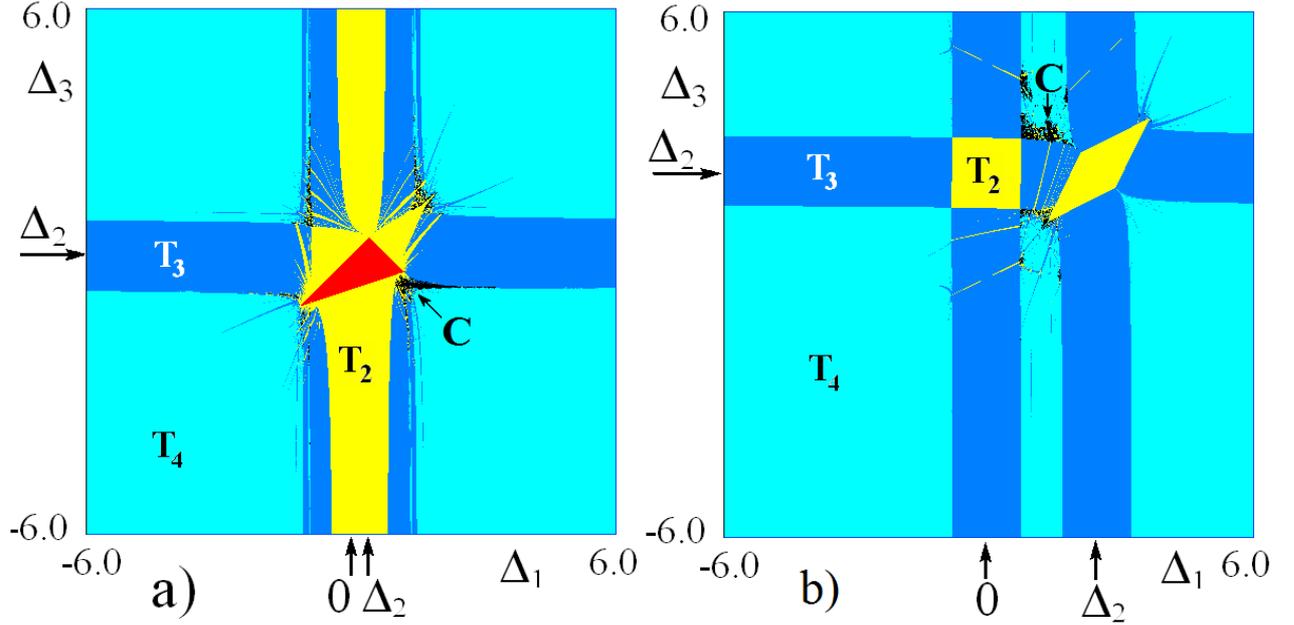

**Fig. 8.** Charts of the Lyapunov exponents for the four dissipatively coupled phase oscillators on the frequency detunings parameter plane $(\Delta_1, \Delta_3)$. Values of the parameters are $\mu = 0.4$, (a) $\Delta_2 = 0.4$, (b) $\Delta_2 = 2.4$. Resonance conditions in the chain of oscillators are shown by arrows.

Possible resonance conditions are shown in Fig. 8a near the bottom and the left boundaries of the chart of Lyapunov exponents. Let us discuss them in detail. There are two possible resonances while varying frequency of the second oscillator $\Delta_1$ on the horizontal axis. This frequency may be equal to those of the first or the third oscillators:

$$\Delta_1 = 0, \quad \Delta_1 = \Delta_2. \tag{13}$$

These resonance points are closely spaced when $\Delta_2$ is small. Presence of two resonances reduces dimension of quasi-periodic regime to two. As a result, fairly wide vertical bar of two-frequency regimes occurs near the area of $\Delta_1 \approx 0 \approx \Delta_2$ on the chart of Lyapunov exponents in Fig. 8a.

During the variation of the forth oscillator's frequency $\Delta_3$ on ordinate axis, the only resonance is possible, because the forth oscillator is on the edge of the chain of oscillators. Resonance occurs when its frequency is equal to the frequency of the third oscillator:

$$\Delta_3 = \Delta_2. \tag{14}$$

So a horizontal bar occurs in the vicinity of the resonance point $\Delta_3 = \Delta_2$ in Fig. 8a, and this is a bar of three-frequency regimes. Full synchronization area is located at the intersection of vertical and horizontal resonance bars.

Now increase the parameter $\Delta_2$. Full synchronization area maintains the shape of a triangle in this case, but reduces in size. The lower side of this triangle reaches its vertex and the full synchronization area completely disappears when $\Delta_2 = 2\mu$. Fig. 8b shows substantially over the threshold case when $\Delta_2 = 2.4$. We can see on the chart of Lyapunov exponents that resonance points (13) diverge along the horizontal axis due to increasing of frequency $\Delta_2$. Situation of resonance overlapping destroys. Bar of two-frequency regimes disappears and is replaced by two vertical bars of three-frequency regimes. Accordingly, full synchronization area disappears. Intersection of bars of three-frequency regimes provides now two regions of two-frequency regimes having the shape of a square and a parallelogram. It is easy to show by means of the phase portraits that inside the square region the first and the second oscillators as well as the third and the fourth oscillators are mutually (partially) captured in pairs. Inside the parallelogram region the second, the third and the forth oscillators are partially captured. We can see also thin stripes corresponding to the two-frequency higher order regimes.

It is interesting to note that the two-frequency area having the shape of a parallelogram in Fig. 8b is qualitatively similar to the full synchronization area of three oscillators named "Landa's parallelogram" in Fig. 1. We can see corresponding fan-shaped system of synchronization tongues in the neighborhoods of the vertices of a parallelogram. This picture is similar to Fig. 1a, but periodic regime is replaced by two-frequency regime and two-frequency regimes are replaced by three-frequency regimes. Finally, note that there are small areas of chaos *C* for the four oscillators in Fig. 8.

Figure 9a shows the chart of Lyapunov exponents for the original system (7) of four oscillators. Values of the parameters are $\lambda = 0.1$, $\mu = 0.04$, $\Delta_2 = 0.02$. Note that parameters $\mu$ and $\Delta_{1,2,3}$ in Fig. 9a are diminished by the factor of $1/\lambda = 10$ as compared to the case of the phase approximation in Fig. 7a. This is due to the rules of normalization to the value of the control parameter [14]. Because this parameter $\lambda$ is not large, the observed picture in Fig. 9a is close to that in Fig. 7a constructed for the phase model. However, there are some differences. For example, shape of the full synchronization area is distorted, periodic higher order resonance regimes appear, etc.

Figure 9b shows the chart of Lyapunov exponents for the case of large value of the control parameter $\lambda = 1$. Now the differences are more significant. The most interesting is the emergence of hierarchically organized resonance web. In this case, a system of bars of three-frequency regimes occurs, which is immersed in the area of four-frequency tori. At the same time thin strips of two-frequency regimes are visible within the bars of three-frequency regimes. This hierarchical organization of the web is the cause of the increased number of oscillators in the system and is different from the case of three oscillators in Fig. 5.

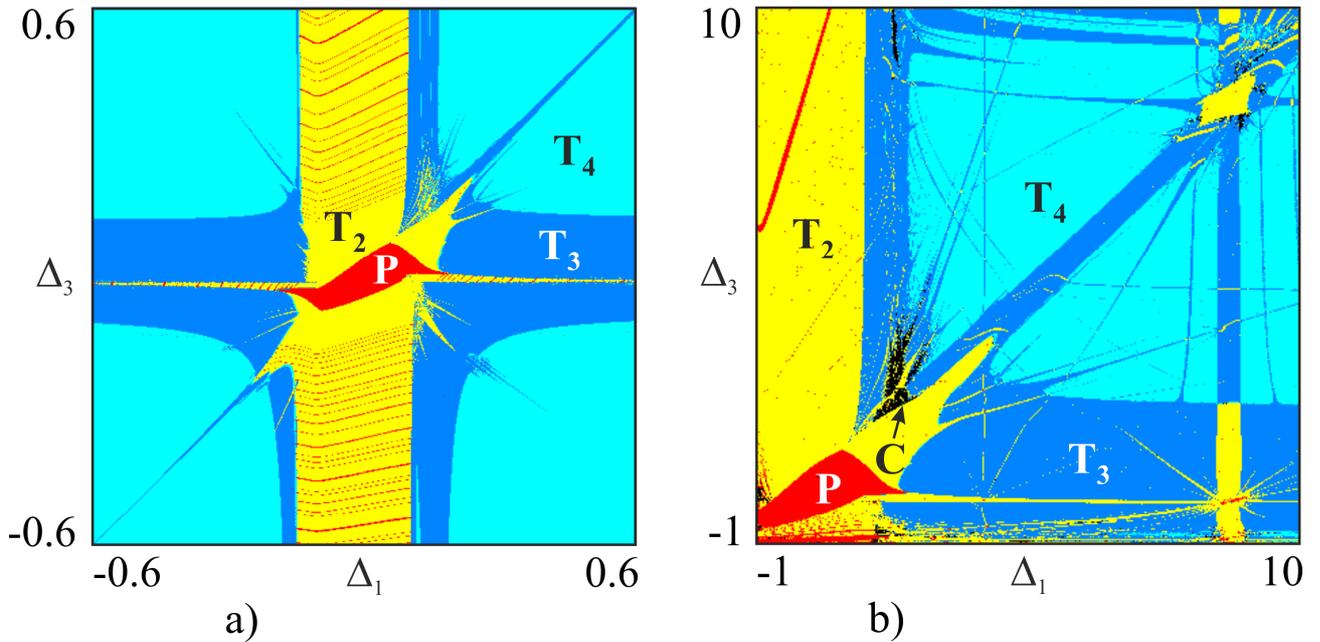

**Fig. 9.** Charts of the Lyapunov exponents for the four coupled van der Pol oscillators (7). Values of the parameters are a) $\lambda = 0.1$, $\mu = 0.04$, $\Delta_2 = 0.02$; b) $\lambda = 1$, $\mu = 0.4$, $\Delta_2 = 0.2$.

## 5. Conclusion

Full synchronization area for the phase model of three dissipatively coupled oscillators looks like a parallelogram on the eigenfrequencies parameter plane. Sides of this parallelogram correspond to the saddle-node bifurcations of such type when both stable and unstable nodes merge simultaneously with saddles. Neighborhoods of vertices of the parallelogram have the most complicated structure. There are fan-shaped systems of two-frequency resonance regimes of different order. Degeneracy disappears for the original system, and the saddle-node bifurcation line splits into the two lines: for the stable and unstable limit cycles. Finite in size areas occur instead of vertices of a parallelogram. Each area contains three cusp points. Neimark–Sacker bifurcations become also possible. Characteristic structure of the Arnol'd resonance web becomes visible for large values of the control parameter. This is a system of thin lines of two-frequency regimes, on intersection of which the higher order periodic resonances arise. In the phase approximation for the system of four coupled oscillators, the full synchronization area looks like an oblique parallelepiped. Its sections may have a form of hexagon or triangle, depending on the fixed frequency of one of the oscillators. If the section does not intersect a polyhedron of the full synchronization, two-frequency quasi-periodic areas may occur. These areas are similar in structure to the case of three oscillators, but the "fan-shaped" structures of synchronization tongues refer now to the three-frequency regimes, immersed in the four-frequency synchronization area.

**Acknowledgments**

The work was supported by RFBR–DFG grant No 11-02-91334.